\documentclass[preprint,epsfig,aps,showpacs]{revtex4}
\usepackage{amsmath}
\usepackage{graphicx}

\newcommand{\f}{\begin{equation}}
\newcommand{\ff}{\end{equation}}
\newcommand{\fa}{\begin{eqnarray}}
\newcommand{\ffa}{\end{eqnarray}}

\begin{document}
\vskip 2.3cm

\title{Extended complex scalar field as quintessence}
\author{Yi-Huan Wei${ }^{1,2}$}
\author{Yuan-Zhong Zhang${ }^{1}$}
\affiliation{%
${ }^1$ Institute of Theoretical Physics, Chinese Academy of
Science, Beijing 100080, China}
\affiliation{%
$ { }^2$  Department of Physics, Bohai University, Jinzhou 121000,
 Liaoning, China}
\begin{abstract}
In this paper, we show the possibility that an extended complex
scalar field can be considered as an extended complex
quintessence. For this model, we derive the basic equations with a
parameter $\eta $ which govern the evolution of the Universe. Our
model may contain a complex quintessence model for $\eta =-1$ and
a real quitessence model for $\eta =0$.

\end{abstract}

\maketitle

\section{Introduction}

The recent astronomy observations on type Ia supernovae suggests
the expansion of the Universe at the present is accelerating
\cite{PR} and measurement of the cosmic microwave background (CMB)
indicate that it is spatially flat \cite{Ber,Han}. Our Universe
has a critical energy density which consists of 1/3 the matter
density and 2/3 of the dark energy density with negative pressure
\cite{Bal}. One proper candidate for the dark energy is a
slowly-evolving real scalar field $\Phi $ with a potential $V$,
called quintessence \cite{Cal,Zlat}. In this model the energy
density and pressure from the scalar quintessence are $\rho _\Phi
=\frac 12\dot{\Phi}^2+V$ and $ P_\Phi =\frac 12\dot{\Phi}^2-V$ and
the state equation is $P_\Phi =w_\Phi \rho _\Phi $ with \footnote{
This upper limit of $w_\Phi $ is determined by following
considerations: providing that for the matter and dark energy
components there are the effective equations of state $P_M=w_M$
$\rho _M$ and $P_\Phi =w_\Phi $ $\rho _\Phi $; from (\ref{7}) to
be shown, for the Universe having an accelerated expansion there
is $3P< -\rho $, which yields $w_M+2w_\Phi < -1$, considering that
$w_M$ is nonnegative we have $w_\Phi < -\frac 12$.} $-1< w_\Phi $
$< -0.5$, the recent study based on the final Cosmic Lens All Sky
Survey data suggest a constraint $w_\Phi $ $< -0.55$ \cite {CBB}.
Recently, the quintessence has extended to the complex scalar
field \cite{Gu,Gao}, $\Phi =\phi e^{i\theta }$, the energy density
and pressure being $\rho _\Phi =\frac 12(\dot{\phi}^2+\phi
^2\dot{\theta}^2)+V,$ $P_\Phi =\frac 12(\dot{\phi}^2+\phi
^2\dot{\theta}^2)-V$. Compared with real quintessence model,
complex quintessence possess the extra term of the energy density
and pressure, $\frac 12\phi ^2\dot{\theta}^2$, the contributions
from the angular-motion. Similarly to the real quintessence model,
the field and potential in the complex quintessence model may be
reconstructed from the observational data by using Huterer and
Turner's procedure \cite{HT}.

In this paper, we will consider an extended complex scalar field
as a quintessence \cite{Wei}. We do so by writing scalar field as
$\Phi _\eta =\phi e^{i_\eta \theta }$, introducing a real
parameter $\eta $ into the energy density and pressure, $\rho
_\Phi =\frac 12(\dot{\phi}^2-\eta
\phi ^2\dot{\theta}^2)+V,$ $P_\Phi =\frac 12(\dot{\phi}^2-\eta \phi ^2\dot{%
\theta}^2)-V$ which for $\eta =-1$ and $\eta =0$ gives those of
complex and real quintessence. When $\eta =1$, our model shows a
new quintessence configuration. As will be shown, in the case of
extended complex quintessence, all quantities including the
potential $V$, the amplitude quintessence field $\phi $, the
equation of state $P_\Phi =w_\Phi \rho _\Phi $ will be
reconstructed in terms of the observable quantities.

\section{Extended complex scalar field as quintessence}

Consider the Robertson-Walker Universe which is spatially flat
\begin{eqnarray}
ds^2=dt^2-a^2(t)(dr^2+r^2d\theta ^2+r^2\sin ^2\theta d\varphi ^2)
\label{1}
\end{eqnarray}
with the action
\begin{eqnarray}
s=\int d^4x\sqrt{-g}[-\frac 1{16\pi G}R-\rho _m+L_\Phi ],
\label{2}
\end{eqnarray}
where $g$ is the matrix determinant of the metric (\ref{1}), $R$
is the Ricci scalar, $G$ is the Newton's constant, $\rho _m$ is
the matter density, and $L_\Phi $ is the Lagrangian density for
extended complex scalar field $\Phi _\eta $ \cite{Wei}
\begin{eqnarray}
L_\Phi =\frac 12\partial ^\mu \Phi _\eta ^{*}\partial _\mu \Phi
_\eta -V(\Phi _\eta ^{*}\Phi _\eta ),  \label{3}
\end{eqnarray}
which is formally the same as that \cite{Gu}, where $\Phi _\eta $
is the extended complex scalar field of the form $\Phi _\eta =Re
\Phi _\eta +i_\eta Im \Phi _\eta $ with $i_\eta $ satisfying
$i_\eta ^2=\eta $ being imaginary unit, $Re\Phi _\eta $ and
$Im\Phi _\eta $ being real and imaginary parts, respectively, and
$\Phi _\eta ^{*}$ $=Re\Phi _\eta +i_\eta Im\Phi _\eta $ is called
the complex conjugate of $\Phi _\eta $ \cite{Wei}. It is easily
proven that the Lagrangian (\ref{3}) possesses the $U(1;i_\eta )$
symmetry\footnote{Considering the following transformation $\Phi
_\eta \rightarrow \Phi _\eta ^{\prime }=e^{i_\eta \alpha }\Phi
_\eta $with $\alpha $ being a real constant and $\Phi _\eta
^{\prime *}=\Phi _\eta ^{*}e^{-i_\eta \alpha }$, then we have
$\Phi _\eta ^{\prime *}\Phi _\eta ^{\prime }=\Phi _\eta
^{*}e^{-i_\eta \alpha }e^{i_\eta \alpha }\Phi _\eta =\phi ^2$ and
$\partial ^\mu \Phi _\eta ^{\prime *}\partial _\mu \Phi _\eta
^{\prime }=\partial ^\mu \Phi _\eta ^{*}\partial _\mu \Phi _\eta $
and therefore $L_\Phi =L_{\Phi ^{\prime }}$. This is called the
$U(1;i_\eta )$ symmetry, which reduces to the $U(1)$ symmetry when
$i_\eta =i$.}.

In contrast to complex scalar field, $\Phi _\eta $ has exponential
form
\begin{eqnarray}
\Phi _\eta =\phi e^{i_\eta \theta }=c(\theta ;i_\eta )+i_\eta
s(\theta ;i_\eta ),  \label{4}
\end{eqnarray}
where $\phi =\sqrt{(Re\Phi _\eta )^2-\eta (Im\Phi _\eta )^2}$ and
$\theta $, a real parameter, is called also ''the amplitude'' and
''phase angular'' of $\Phi _\eta $, respectively. Putting
(\ref{4}) into the Lagrangian (\ref{3}) we have
\begin{eqnarray}
L_\Phi =\frac 12(\partial ^\mu \phi \partial _\mu \phi -\eta \phi
^2\partial ^\mu \theta \partial _\mu \theta )-V(\phi ),  \label{5}
\end{eqnarray}
which becomes the Lagrangian density (5) in Ref.\cite{Gu} for
$\eta =-1$ and gives the following Lagrangian density for $\eta
=1$
\begin{eqnarray}
L_\Phi =\frac 12(\partial ^\mu \phi \partial _\mu \phi -\phi
^2\partial ^\mu \theta \partial _\mu \theta )-V(\phi ). \label{5a}
\end{eqnarray}

The variation of the action (1) with the Lagrangian density
(\ref{5}) leads to the equations of motion
\begin{eqnarray}
H^2=\frac{8\pi G}3\rho ,  \label{6}
\end{eqnarray}
\begin{eqnarray}
\frac{\ddot{a}}a=-\frac{8\pi G}3(\rho +3P),  \label{7}
\end{eqnarray}
\begin{eqnarray}
\ddot{\phi}+3H\dot{\phi}+\eta \phi \dot{\theta}^2+V^{\prime }=0,
\label{8}
\end{eqnarray}
\begin{eqnarray}
\ddot{\theta}+(2\frac{\dot{\phi}}\phi +3H)\dot{\theta}=0,
\label{9}
\end{eqnarray}
where a dot and prime denotes derivatives with respect to $t$ and
$\phi $, respectively, $H=\dot{a}/a$ is the Hubble parameter,
$\rho $ and $P$ are the total density and pressure of the Universe
\begin{eqnarray}
\rho =\rho _m+\rho _\Phi ,\quad \rho _\Phi =\frac
12(\dot{\phi}^2-\eta \phi ^2\dot{\theta}^2)+V,  \label{10}
\end{eqnarray}
\begin{eqnarray}
P=P_m+P_\Phi ,\quad P_\Phi =\frac 12(\dot{\phi}^2-\eta \phi
^2\dot{\theta} ^2)-V  \label{11}
\end{eqnarray}
with $\rho _\Phi $ and $P_\Phi $ denote the energy density and
pressure from the scalar field $\Phi _\eta $ and potential $V$,
called an extended complex scalar quintessence. $\rho _\Phi $ and
$P_\Phi $ in (\ref{10}) and ( \ref{11}) consist of the three
terms: the kinetic energy contribution from the amplitude of the
extended complex scalar field $\frac 12\dot{\phi}^2$, the 'kinetic
energy' contribution from the 'angular-motion' $\frac 12\eta \phi
^2\dot{\theta}^2$ and the potential energy $V$.

Note that equation (\ref{9}) is independent of the parameter $\eta
$, for arbitrary $\eta $ there is the solution
\begin{eqnarray}
\dot{\theta}=ca^{-3}\phi ^{-2},  \label{12a}
\end{eqnarray}
where $c$ is an integration constant. According to \cite{Gu}, the
term $\phi ^2\dot{\theta}^2$ from the angular-motion in the
quintessence energy density $\rho _\Phi $ and pressure $P_\Phi $
may be treated as an effective potential. Clearly, such an
interpretation for the term $\phi ^2\dot{\theta} ^2$ is improper
since $\rho _\Phi $ and $P_\Phi $ takes the form $\rho _\Phi
=\frac 12\dot{\phi}^2+(V-\eta \phi ^2\dot{\theta}^2)$ and $P_\Phi
=\frac 12 \dot{\phi}^2-(V+\eta \phi ^2\dot{\theta}^2)$. It is more
like an effective kinetic energy than a potential energy.
Certainly, if the term $\frac 12\eta \phi ^2\dot{\theta}^2$ is
treated as effective kinetic for $\eta =1$ it will take an
negative value, this seem to be unreasonable. For this, the
further discussion will be given in Sec.3. Now, let's turn our
attention to the problem of reconstruction equations.

From the relationships
\begin{eqnarray}
\begin{array}{l}
a=\frac {1}{z+1}, \\
r(z)=\int_0^z\frac{dz^{\prime }}{H(z^{\prime })}, \\
H(z)=\frac{\dot{a}}a=\frac 1{(dr/dz)}, \\
\rho _M=\Omega _M\rho _c=\frac{3\Omega _MH_0^2}{16\pi G}(1+z)^3,
\end{array}
\label{a}
\end{eqnarray}
where $z$ $\rho _c$ $\Omega _M,$ $H_0$ and $r(z)$ are the
redshift, the critical density, the matter energy density fraction
and the coordinate distance to an object at redshift $z$, the
reconstruction equations of the potential function $V$, the
amplitude $\phi $ and phase angular of extended complex
quintessence in terms of the observational quantities are
\begin{eqnarray}
V=\frac 1{8\pi G}[\frac
3{(dr/dz)^2}+(1+z)\frac{d^2r/dz^2}{(dr/dz)^3}]-\frac{ 3\Omega
_MH_0^2}{16\pi G}(1+z)^3,  \label{13}
\end{eqnarray}
\begin{eqnarray}
(\frac{d\phi }{dz})^2-\eta \frac{c^2}{\phi
^2}(1+z)^4(\frac{dr}{dz})^2=\frac{ (dr/dz)^2}{(1+z)^2}[-\frac
1{4\pi G}\frac{(1+z)d^2r/dz^2}{(dr/dz)^3}-\frac{ 3\Omega
_MH_0^2}{8\pi G}(1+z)^3],  \label{14}
\end{eqnarray}
\begin{eqnarray}
\frac{d\theta }{dz}=-c(1+z)^2H^{-1}(z)\phi ^{-2}  \label{Ang}
\end{eqnarray}
where $\dot{\phi}=\frac{d\phi }{dz}\frac{dz}{dt}=-\frac{d\phi
}{dz}\frac{ \dot{a}}{a^2}$ has been used in deriving equation
(\ref{14}). For $\eta =-1$, (\ref{14}) give equations (25) in Ref.
\cite{Gu}, and for $\eta =0$ or $c=0$ it gives the real scalar
field equation \cite{HT}. Interestingly, for either the real,
complex or extended complex scalar quintessence model, the
reconstruction potential $V$ is the same one. For the
angular-motion contribution being negligible \cite{Gu} (compared
with the evolving-$\phi $), there is
\begin{eqnarray}
(\frac{d\phi }{dz})^2=\frac{(dr/dz)^2}{(1+z)^2}[-\frac 1{4\pi
G}\frac{ (1+z)d^2r/dz^2}{(dr/dz)^3}-\frac{3\Omega _MH_0^2}{8\pi
G}(1+z)^3], \label{16}
\end{eqnarray}
and for the evolving-$\phi $ contribution being negligible
(compared with the angular-motion contribution) there is
\begin{eqnarray}
\eta \frac{c^2}{\phi ^2}=\frac 1{(1+z)^6}[\frac 1{4\pi
G}\frac{(1+z)d^2r/dz^2 }{(dr/dz)^3}+\frac{3\Omega _MH_0^2}{8\pi
G}(1+z)^3].  \label{18}
\end{eqnarray}

Noting that
\begin{eqnarray}
\dot{\phi}=-\frac{d\phi
}{dz}\frac{\dot{a}}{a^2}=-\frac{(1+z)}{(dr/dz)}\frac{ d\phi }{dz},
\label{19}
\end{eqnarray}
for the extended complex quintessence we obtain the pressure of
the energy density given in terms of $z$ as
\begin{eqnarray}
P_\Phi =-\frac 1{8\pi G}[\frac
3{(dr/dz)^2}+2(1+z)\frac{d^2r/dz^2}{(dr/dz)^3} ],  \label{20}
\end{eqnarray}
\begin{eqnarray}
\rho _\Phi =\frac 3{8\pi G}[\frac 1{(dr/dz)^2}-\Omega
_MH_0^2(1+z)^3], \label{21}
\end{eqnarray}
and the reconstruction equation of the state
\begin{eqnarray}
w_\Phi =\frac 13\frac{2(1+z)(d^2r/dz^2)+3(dr/dz)}{\Omega
_MH_0^2(1+z)^3(dr/dz)^3-(dr/dz)}.  \label{22}
\end{eqnarray}
This reconstruction equation (\ref{22}) is the same one as (17) in
Ref.\cite {HT}.

\section{Conclusions}

For non-minimally coupling to the curvature, the Lagrangian
density is
\begin{eqnarray}
L_\Phi =\frac 12\partial ^\mu \Phi _\eta ^{*}\partial _\mu \Phi
_\eta -V(\Phi _\eta ^{*}\Phi _\eta )-\frac 12\xi R\Phi _\eta
^{*}\Phi _\eta , \label{23}
\end{eqnarray}
for $i_\eta =i$ which gives the one \cite{Gao}. In this case, the
energy density and pressure are
\begin{eqnarray}
\rho _\Phi =\frac 12(\dot{\phi}^2-\eta \phi
^2\dot{\theta}^2)+V-\frac 32\xi ^2R\phi ^2+\xi (-\frac 32\phi
\ddot{\phi}+\frac 32H\phi \dot{\phi}+3H^2\phi ^2-3V),  \label{24}
\end{eqnarray}
\begin{eqnarray}
P_\Phi =\frac 12(\dot{\phi}^2-\eta \phi ^2\dot{\theta}^2)-V+\frac
32\xi ^2R\phi ^2-\xi (2\dot{\phi}^2+\frac 12\phi \ddot{\phi}+\frac
32H\phi \dot{ \phi}-\phi ^2G_1^1-3V),  \label{25}
\end{eqnarray}
which may be reformulated as
\begin{eqnarray}
\rho _\Phi =E_{eff}+V_{eff},  \label{26}
\end{eqnarray}
\begin{eqnarray}
P_\Phi =E_{eff}-V_{eff}  \label{27}
\end{eqnarray}
with
\begin{eqnarray}
E_{eff}=\frac 12(\dot{\phi}^2-\eta \phi ^2\dot{\theta}^2)+\frac
12\xi (-2\dot{\phi}^2-2\phi \ddot{\phi}+6H^2\phi ^2),  \label{28}
\end{eqnarray}
\begin{eqnarray}
V_{eff}=V-\frac 32\xi ^2R\phi ^2+\frac 12\xi (2\dot{\phi}^2-\phi
\ddot{\phi}+3H\phi \dot{\phi}-6V).  \label{29}
\end{eqnarray}
Obviously, for the Lagrangian density, there always are the
reconstruction equations
\begin{eqnarray}
V_{eff}=\frac 1{8\pi G}[\frac
3{(dr/dz)^2}+(1+z)\frac{d^2r/dz^2}{(dr/dz)^3}]-\frac{3\Omega
_MH_0^2}{16\pi G}(1+z)^3,  \label{30}
\end{eqnarray}
\begin{eqnarray}
E_{eff}=-\frac{(1+z)}{4\pi
G}[\frac{d^2r/dz^2}{(dr/dz)^3}-\frac{3\Omega _MH_0^2}2(1+z)^2],
\label{31}
\end{eqnarray}
In fact, the form of the energy density and pressure (\ref{26})
and (\ref{27}) are general and may applied to any model
interpreted as dark energy, so does the reconstruction equations
for $E_{eff}$ and $V_{eff}$.

In different model $E_{eff}$ and $V_{eff}$ take different form
means different understanding of the origin of the dark energy. We
propose the extended complex quintessence model, in which a real
parameter is introduced. For $\eta =-1$, it gives the complex
quintessence case. As discussed above, term $\frac 12\eta \phi
^2\dot{\theta}^2$ appearing in the energy density and the pressure
can be treated as an effective kinetic energy. However, for $\eta
=1$ it contributes a negative kinetic energy. Here, we would like
to consider the Lagrangian density $L_\Phi $ as a sum of the two
parts: $L_\Phi =L_\phi +L_{a\phi }$ with $L_\phi =\frac 12\partial
^\mu \phi \partial _\mu \phi -V(\phi )$ and $L_{a\phi }=-\frac
12\eta c^2a^{-6}\phi ^{-2}$. The first term is equivalent to the
Lagrangian density of a real scalar field and the second term may
be explained as a Lagrangian density from the coupling of $\phi$
to $a$, noting $\dot{\theta}=ca^{-3}\phi ^{-2}$. The Lagrangian
density $L_{a\phi }$ contributes the equivalent pressure and
energy density in the Universe $P_c=\rho _c=-\frac 12\eta
c^2a^{-6}\phi ^{-2}$ and may be reconstructed as
\begin{eqnarray}
L_{a\phi }=-\frac 1{8\pi
G}\frac{(1+z)d^2r/dz^2}{(dr/dz)^3}-\frac{3\Omega
_MH_0^2(1+z)^3}{16\pi G}-(1+z)^6(\frac{d\phi }{dr})^2. \label{Lc1}
\end{eqnarray}

Compared to the previous scalar quintessence model, the extended
complex quintessence model has the following characteristic. The
so-called slowly-evolving \cite{Zlat} scalar field in the complex
or real quintessence model implies only $\dot{\phi}^2-\phi
^2\dot{\theta}^2$ is slowly evolving, here, so, the field $\phi$
could be possibly moving with the kinetic energy over the
potential energy, even it may be fastly evolving since $\phi
^2\dot{ \theta}^2$ is positive term and hasn't been determined.
Another point is, as discussed above, the extended complex
quintessence model may applied to the future observational data in
a wider range. In addition, for $\eta =1$ case, there is the
possibility of the equation of the state with $w_\Phi < -1$ if the
condition $\dot{\phi}^2-\phi ^2\dot{\theta}^2< 0$ is satisfied.
Although the observation data at the present implies the equation
of state for dark energy component $w_\Phi =P_\Phi /\rho _\Phi
\leq -1$, it must not be the constraint condition on the future
Universe since we cannot specify how the Universe has evolved or
will evolve.

It is also an important and interesting fact that the
reconstruction equation of the potential $V$ doesn't depend on the
scalar quintessence models (real, complex, or extended complex),
namely, it is independent of the parameter $\eta $. This feature
of the scalar quintessence means that our generalization to the
work \cite{HT,Gu,Gao} should be natural.

\end{document}